\documentclass[pra,aps,preprint,doublespace]{revtex4}
\usepackage{amssymb}
\usepackage{amsmath}
\usepackage{epsfig}

\begin{document}

\title{Efficient fiber coupling of down-conversion photon pairs}

\date{\today}

\author{Andrzej Dragan}
\affiliation{Instytut Fizyki Teoretycznej, Uniwersytet Warszawski,
Ho\.{z}a 69, PL-00-681 Warszawa, Poland}

\begin{abstract}
We develop and apply an effective analytic theory of a
non-collinear, broadband type-I parametric down-conversion to
study a coupling efficiency of the generated photon pairs into
single mode optical fibers. We derive conditions necessary for
highly efficient coupling for single and double type-I crystal
producing polarization entangled states of light. We compare the
obtained approximate analytic expressions with the exact numerical
solutions and discuss the results for a case of BBO crystals.
\end{abstract}

\maketitle

\section{Introduction}
Sources of polarization entangled photon pairs are important in a
wide range of modern applications. Efficient generation of photon
pairs led to an opportunity of testing several fundamental
problems of quantum mechanics and to a practical realization of
various quantum-information or cryptographic schemes. Among all
accessible entanglement sources, down-conversion crystals turned
out to be the most effective and easy to implement so far.
Therefore the development of down-conversion techniques started to
be an important issue for many branches of quantum optics
\cite{ref-Kwiat1995}. In a search for enhancement of counting
rates, it has been noticed that focusing of the pump beam can
essentially increase the efficiency of the process
\cite{ref-Monken1998}. Another important observation was that
coupling of the generated photons into single mode fibers can
optimize the collection efficiency and is a very practical method
of preparing photons in well defined spatial modes
\cite{ref-Kurtsiefer2001}. This idea has already inspired some
authors to study various aspects of fiber coupling of the
down-converted photon pairs
\cite{ref-Bovino2003,ref-Castalletto2003}. Efficient fiber
coupling of the entangled photon pairs can also help in enhancing
the fiber communication efficiency by encoding additional
information in the polarization degree of freedom
\cite{ref-Banaszek2004}.

The paper delivers a detailed analysis of the fiber coupling
efficiency and properties of the coupled type-I down-conversion
photon pairs in a general case of a pumping with ultrashort
pulses. To our best knowledge this issue has not been analyzed so
far. One of the advantages of using the pulsed pump sources is the
fact, that generated photon pairs arrive at the detectors at a
well-defined temporal window which may be very helpful in a number
of experimental schemes \cite{ref-Banaszek2004}.

We develop an approximate analytic theory of type-I
down-conversion to study the dependence of the coincidence
spectrum and coupling efficiency on several relevant parameters
characterizing the system such as properties of the nonlinear
medium, pumping beam, spectral filters, geometry of the setup,
etc. We point out what are the optimum (for the coupling
efficiency and a spectrum separability, which is necessary for a
high-purity polarization entanglement generation
\cite{ref-URen2003}) settings of the setup and compare the result
with the numerical calculations.

The paper is organized as follows: in Sec. \ref{sec-WaveFunction}
we derive explicitly a wave function of the type-I down-conversion
generated photon pairs and discuss the applied approximations. In
Sec. \ref{sec-Coupling} we discuss the coupling process,
properties of the spectra of the coupled photons, derive the
spectrum separability condition and derive expressions for the
coupling probability. In Sec. \ref{sec-Double} we apply the
obtained results to discuss the case of a double type-I crystal
generating polarization entangled states and finally Sec.
\ref{sec-Conclusions} concludes the paper and in the Appendix we
derive a relation between geometrical parameters of the crystal
and a dispersion relation.

\section{Wave function\label{sec-WaveFunction}}
Consider a type-I down-conversion process taking place in an
anisotropic medium with a second-order nonlinearity. Interaction
between an extraordinarily polarized laser pump beam
$E_{\mbox{\scriptsize P}}({\bf r},t)$ (which will be assumed to
have a finite time) and the nonlinear crystal generates pairs of
daughter photons whose properties are fully determined by the
properties of the pump beam, indices of refraction and geometry of
the process - see Figure \ref{fig-schemat}. In the first order
perturbation, only a single photon pair can be created in the
down-conversion and the output quantum state has the approximate
form:

\begin{equation}
\label{eq-State01} |\Psi\rangle \approx |0\rangle
-i\frac{\chi}{\hbar}\int_{-\infty}^\infty \mbox{d}t\int_V
\mbox{d}^3{\bf r}\,E_{\mbox{\scriptsize P}}({\bf r},t)
\left(\hat{E}^{(-)}({\bf r},t)\right)^2|0\rangle,
\end{equation}
where an electric nonlinear polarizability $\chi$ is a parameter
determining the interaction strength, $V$ is volume of the crystal
and a negative part of the electric field operator is
$\hat{E}^{(-)}({\bf
r},t)=\int\mbox{d}k_x\,\mbox{d}k_y\,\mbox{d}\omega\, e^{-i({\bf
k}\cdot{\bf r}-\omega t)}\hat{a}^\dagger(k_x,k_y,\omega)$
($\hat{a}$ is an annihilation operator of an ordinarily polarized
mode, at $z=0$). The integral extends to all positive frequencies
$\omega$ and all perpendicular wave-vectors $k_x$, $k_y$. Such a
choice of the plane wave decomposition will turn out to be very
handy, since the perpendicular wave-vector components and
frequency do not change on dielectric media boundaries
\cite{ref-Keller1997,ref-Rubin1997}. Let us introduce signal and
idler mode vectors ${\bf s} = (k_{\mbox{\scriptsize
s},x},k_{\mbox{\scriptsize s},y}, \omega_{\mbox{\scriptsize s}})$
and ${\bf i} = (k_{\mbox{\scriptsize i},x},k_{\mbox{\scriptsize
i},y}, \omega_{\mbox{\scriptsize i}})$ to decompose the square of
the electric field operator appearing in the expression
(\ref{eq-State01}). Then the $z$th components of the wave vectors
are given by a dispersion relations inside the anisotropic medium
$k_{\mbox{\scriptsize s},z} =k^{\mbox{\scriptsize o}}_z({\bf s})$
and $k_{\mbox{\scriptsize i},z} =k^{\mbox{\scriptsize o}}_z({\bf
i})$ (index "o" refers to ordinary wave of the considered
anisotropic medium - see Appendix) and the output quantum state
can be written down using a quantity $\Psi({\bf s},{\bf i})$
interpreted as a probability amplitude for a photon pair to be
created in modes ${\bf s}$ and ${\bf i}$:

\begin{equation}
\label{eq-State02} |\Psi\rangle \approx |0\rangle
-i\frac{\chi}{\hbar}\int\mbox{d}^3{\bf s}\,\mbox{d}^3{\bf
i}\,\Psi({\bf s},{\bf i})\hat{a}^\dagger({\bf
s})\hat{a}^\dagger({\bf i})|0\rangle.
\end{equation}
The first order approximation that we will study is justified when
$\chi\ll 1$. The properties of the generated photon pairs are
described by the function $\Psi({\bf s},{\bf i})$ and this is a
typical situation in many down-conversion experiments.

Suppose that a type-I down-conversion medium occupies the area
$-\frac{L}{2}\leq z\leq \frac{L}{2}$ and it is pumped by a light
beam propagating towards $z$ direction. Then the probability
amplitude $\Psi({\bf s},{\bf i})$ at the end of the crystal has
the following form:

\begin{eqnarray}
\label{eq-PsiAnalytic0} \Psi({\bf s},{\bf i}) &=&
\int_{-\infty}^\infty \mbox{d}t\int_{-\infty}^\infty
\mbox{d}x\,\int_{-\infty}^\infty \mbox{d}y\,\int_{-z_1}^{z_2}
\mbox{d}z\,E_{\mbox{\scriptsize P}}({\bf r},t)\times
\nonumber \\ \nonumber \\
& &e^{-i\left[({\bf k}_{\mbox{\tiny s}}+{\bf k}_{\mbox{\tiny i}})
\cdot{\bf r}-(\omega_{\mbox{\tiny s}}+\omega_{\mbox{\tiny i}})
t\right]}e^{i(k_{\mbox{\tiny s},z}+k_{\mbox{\tiny
i},z})\frac{L}{2}},
\end{eqnarray}
where the last exponent is a propagation factor of the amplitude
from $z=0$ to $z=\frac{L}{2}$.

Let us express the pump beam electric field in terms of its
Fourier transform defined at $z=-\frac{L}{2}$:
$E_{\mbox{\scriptsize P}}({\bf
r},t)=\int\mbox{d}k_{\mbox{\scriptsize p},x}\,
\mbox{d}k_{\mbox{\scriptsize p},y}\,
\mbox{d}\omega_{\mbox{\scriptsize p}}\, e^{i({\bf k}_{\mbox{\tiny
p}} \cdot{\bf r}-\omega_{\mbox{\tiny p}} t)} e^{ik_{\mbox{\tiny
p},z}\frac{L}{2}} \tilde{E}_{\mbox{\scriptsize
P}}(k_{\mbox{\scriptsize p},x}, k_{\mbox{\scriptsize
p},y},\omega_{\mbox{\scriptsize p}})$, where we have introduced
vector ${\bf p} = (k_{\mbox{\scriptsize p},x},k_{\mbox{\scriptsize
p},y}, \omega_{\mbox{\scriptsize p}})$ parameterizing the plane
wave decomposition of the pump beam. To calculate the probability
amplitude (\ref{eq-PsiAnalytic0}) we need the dispersion relation
$k_{\mbox{\scriptsize p},z} =k^{\mbox{\scriptsize e}}_z({\bf p})$,
which is well known for uniaxial crystals \cite{ref-Yariv} - see
Appendix. Inserting this into the equation (\ref{eq-PsiAnalytic0})
and performing all the integrals yields \cite{ref-Klyshko1988}:

\begin{equation}
\label{eq-PsiAnalytic} \Psi({\bf s }, {\bf i}) =
\tilde{E}_{\mbox{\scriptsize P}}({\bf s}+{\bf i}) L \, \mbox{sinc}
\left(\frac{L}{2}\delta_-({\bf s }, {\bf i})\right)\exp\left(
i\frac{L}{2}\delta_+({\bf s }, {\bf i})\right),
\end{equation}
where $\delta_\pm$ ($\delta_-$ is a so-called phase mismatch
parameter) equal:

\begin{equation}
\label{eq-PhaseMismatchAnalytic} \delta_\pm({\bf s }, {\bf i}) =
k^{\mbox{\scriptsize e}}_z({\bf s }+{\bf i})\pm
k^{\mbox{\scriptsize o}}_z({\bf s })\pm k^{\mbox{\scriptsize o}}_z
({\bf i}).
\end{equation}
From the form of equation (\ref{eq-PsiAnalytic}) it is clear that
the process occurs most effectively in the directions for which
the longitudinal wave vector components meet the phase matching
condition $\delta_-=0$. To determine this condition explicitly one
has to use Sellmeier expressions for the extraordinary and
ordinary index of refraction as a function of the vectors ${\bf s
}$ and ${\bf i}$.

Since the expression (\ref{eq-PsiAnalytic}) is not analytically
integrable, which will turn out to be crucial in our analysis and
the explicit form of the mismatch parameter $\delta_-$ is rather
complicated, we will perform certain approximations before
starting any further analysis. First of all, let us perform a
Gaussian approximation of the $\mbox{sinc}\,x$ function appearing
in (\ref{eq-PsiAnalytic}) with the expression $e^{-x^2/4}$. This
rough approximation is fair as long as we consider only photons
generated near the phase-matching region of wave vectors and
frequencies where the down-conversion occurs most effectively. In
this narrow range of parameters. i.e. when $\frac{L}{2}\delta_-\ll
1$ the wave tails of the $\mbox{sinc}\,x$ function give
insignificant contribution to the overall probability distribution
of generating a pair of photons.

Secondly, we will expand $\delta_\pm$ to the first order in the
Taylor series around  a point for which the phase-matching
condition $\delta_-=0$ is met in the horizontal plane. If we fix
the $x$ axis to be horizontal and the $y$ axis to be vertical, our
expansion takes place around the point ${\bf s}_0 =
\left(\frac{\theta_0\omega_0}{c},0,\omega_0\right)$ and ${\bf i}_0
= \left(-\frac{\theta_0\omega_0}{c},0,\omega_0\right)$, where
$\omega_0$ is the phase-matching degenerate frequency and
$\theta_0$ is the angle of propagation of the degenerate daughter
photons relative to the $z$ axis outside the crystal. These two
parameters can be determined from the properties of the
down-conversion medium (given the Sellmeier formulas for
extraordinary indices of refraction and direction of the crystal
axis, one can calculate $\omega_0$ and $\theta_0$ directly from
the condition $\delta_-({\bf s}_0,{\bf i}_0)=0$ and the definition
(\ref{eq-PhaseMismatchAnalytic})). We obtain an approximate form
of the probability amplitude (\ref{eq-PsiAnalytic}):

\begin{equation}
\label{eq-PsiApproximated} \Psi({\bf s},{\bf i})
\approx\tilde{E}_{\mbox{\scriptsize P}}({\bf s}+{\bf i})
L\exp\left[-\frac{L^2}{16}\left({\cal D}
\delta_-\right)^2+i\frac{L}{2}{\cal D}\delta_+\right],
\end{equation}
where we have introduced the following notation: ${\cal
D}f(x_1,...x_N) = \sum_{i=1}^N\frac{\partial f}{\partial
x_i}\Delta x_i$. It is apparent that in our approximation all the
relevant information about the structure of the nonlinear medium
is contained in the derivatives of $\delta_\pm$. Particular
components of these vectors depend on the properties of both,
extraordinary and ordinary indices of refraction, as well as on
the spatial orientation of the optical axis.

For the reasons that will be made clear in the further parts of
the analysis we shall choose the optical axis of the crystal in a
plane oriented at the angle $45^\circ$ in respect to the
horizontal plane. In this case all the derivatives considered can
be approximately expressed with only 4 real parameters (moreover,
it will turn out that one of them is irrelevant, so we end up with
only 3 parameters) - see Appendix:

\begin{eqnarray}
\label{eq-mismatchderivatives} \frac{\partial\delta_\pm}{\partial
k_{\mbox{\scriptsize s},x}} &=&
\frac{\gamma}{\sqrt{2}} \mp \theta'_0 \nonumber \\
\frac{\partial\delta_\pm}{\partial k_{\mbox{\scriptsize i},x}} &=&
\frac{\gamma}{\sqrt{2}} \pm \theta'_0 \nonumber \\
\frac{\partial\delta_\pm}{\partial k_{\mbox{\scriptsize s},y}} &=&
\frac{\partial\delta_\pm}{\partial k_{\mbox{\scriptsize i},y}} =
\frac{\gamma}{\sqrt{2}} \nonumber \\
\frac{\partial\delta_\pm}{\partial \omega_{\mbox{\scriptsize s}}}
&=& \frac{\partial\delta_\pm}{\partial \omega_{\mbox{\scriptsize
i}}} \approx
\Delta\beta_{\pm,z}, \nonumber \\
\end{eqnarray}
where $\theta'_0$ is the angle of propagation of the degenerate
daughter photons relative to the $z$ axis inside the crystal
proportional to the $\theta_0$ (proportionality constant is given
by the ordinary index of refraction), $\gamma$ (equal to the
derivative of $k_z^{\mbox{\scriptsize e}}$ in the direction of
projection of the optical axis onto the plane transversal respect
to the $z$ axis) is a walk-off angle of the extraordinary ray of
frequency $2\omega_0$ incident in the $z$ direction and
$\Delta\beta_{\pm,z}=\beta^{\mbox{\scriptsize e}}_z(2\omega_0)\pm
\beta^{\mbox{\scriptsize o}}_z(\omega_0)$ are the combinations of
the extraordinary and ordinary inverse group velocities towards
$z$ axis for the frequencies $2\omega_0$ and $\omega_0$,
respectively. We have also assumed that the $z$-th component of
the ordinary inverse group velocity $\beta^{\mbox{\scriptsize
o}}_z(\omega)$ weakly depends on the direction of propagation.

\section{Coupling photon pairs into fibers\label{sec-Coupling}}
As we have seen, the full information about the modal structure of
the generated photon pairs is contained in the function $\Psi({\bf
s},{\bf i})$ - the probability amplitude of emission of a photon
pair of certain frequencies towards given two directions. In the
next step of our analysis, we will study how these photon pairs
are coupled into single-mode optical fibers. We may think of such
fibers as a pair of spatial filters letting through the photons
occupying only selected transverse modes $u(x,y,\omega)$
propagating through the fiber.

\begin{figure}
\begin{center}
\epsfig{file=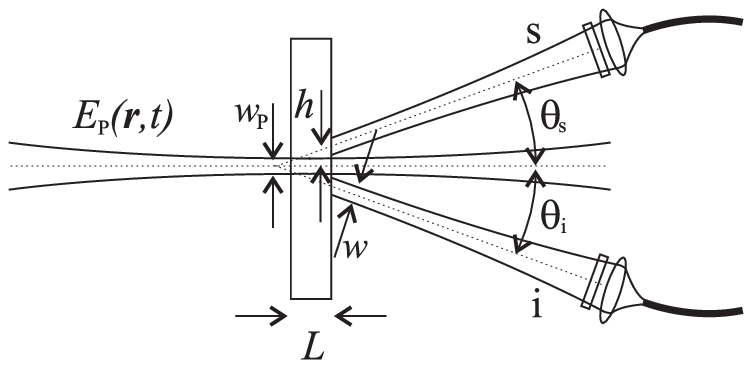}
\end{center}
\caption{\label{fig-schemat} Scheme of the setup: a pump beam
$E_{\mbox{\scriptsize P}}({\bf r},t)$ focused on the nonlinear
crystal (length $L$) with a waist $w_{\mbox{\scriptsize P}}$ and
the coupled modes with waists $w$ and a shift $h$ at the output
face of the crystal in respect to the pump beam. The coupled modes
$\mbox{s}$ and $\mbox{i}$ are oriented at the angles
$\theta_{\mbox{\scriptsize s}}$ and $\theta_{\mbox{\scriptsize
i}}$ with respect to the $z$ axis and the coupling lenses are
preceded with interference filters.}
\end{figure}

Consider a scheme in which photons generated in the crystal pass
through interference filters and lenses coupling incident light
into the single-mode fibers. Let the fibers be oriented in the
horizontal plane at the angles $\theta_{\mbox{\scriptsize s}}$ and
$\theta_{\mbox{\scriptsize i}}$ with respect to the $z$ axis - see
Figure \ref{fig-schemat}. The coupled modes can be easily
characterized if we consider an inverse problem: what is the
structure of a light mode emerging from the fiber? The output mode
$u(x,y,\omega)$ can be approximated by a Gaussian beam incident on
the down-conversion crystal. Let us consider a setup in which the
coupled beam propagating in the horizontal plane (common with the
pump beam) crosses the output face of the crystal of a distance
$h$ from the point of intersection of the pump beam and the
crystal surface - see Figure \ref{fig-schemat}. In the simplest
case, when the output beam is focused on the surface of the
crystal, it is parameterized only by the angle of incidence
$\theta_{\mbox{\scriptsize s}}$, the distance $h$ from the pump
beam and the waist $w$. The Fourier-transformed mode function at
the output surface of the crystal for the signal arm (and
analogously, only with $-h$ instead of $h$ for the idler arm) has
the form:

\begin{widetext}
\begin{equation}
\tilde{u}({\bf s}) \approx A_{\mbox{\scriptsize
F}}(\omega_{\mbox{\scriptsize
s}})\frac{w}{\sqrt{2\pi}}\exp\left\{-\frac{w^2}{4}
\left[\left(k_{\mbox{\scriptsize
s},x}-\frac{\omega_{\mbox{\scriptsize s}}
\theta_{\mbox{\scriptsize s}}}{c}\right)^2+k^2_{\mbox{\scriptsize
s},y} \right] -ih\left(k_{\mbox{\scriptsize s},x}
-\frac{\omega_{\mbox{\scriptsize s}} \theta_{\mbox{\scriptsize
s}}}{c}\right) \right\},
\end{equation}
\end{widetext}
where $A_{\mbox{\scriptsize F}} (\omega) = \exp\left(
-\frac{(\omega-\omega_0)^2} {2\sigma_{\mbox{\tiny F}}^2} \right)$
characterizes transmissivity of the interference filters
characterized by their spectral bandwidth
$\sigma_{\mbox{\scriptsize F}}$. The probability amplitude
$\psi(\omega_{\mbox{\scriptsize s}},\omega_{\mbox{\scriptsize
i}})$ for the photon pair of frequencies
$\omega_{\mbox{\scriptsize s}}$ and $\omega_{\mbox{\scriptsize
i}}$ to be coupled into the fibers reads:

\begin{equation}
\label{eq-PsiFrequencies} \psi(\omega_{\mbox{\scriptsize
s}},\omega_{\mbox{\scriptsize i}}) =
\int\mbox{d}k_{\mbox{\scriptsize s},x}
\mbox{d}k_{\mbox{\scriptsize s},y}\mbox{d}k_{\mbox{\scriptsize
i},x}\mbox{d}k_{\mbox{\scriptsize i},y} \Psi({\bf s},{\bf i})
\tilde{u}^*({\bf s})\tilde{u}^*({\bf i}).
\end{equation}
Notice, that the above expression involves integrating of the
probability amplitudes, not probabilities, because photons with
different wave vectors, but the same frequency are
indistinguishable after coupling into the fibers.

To write down explicitly the amplitude
$\psi(\omega_{\mbox{\scriptsize s}},\omega_{\mbox{\scriptsize
i}})$ it is useful to introduce a set of effective parameters
describing the process: a relative pump beam walk-off parameter
$\Gamma$, a relative shift of the degenerate photons $\Theta$ and
an effective crystal length ${\cal L}$:

\begin{eqnarray}
\label{eq-effective parameters}
\Gamma &=& \frac{L\gamma}{\sqrt{w^2+2w_{\mbox{\scriptsize P}}^2}}\nonumber \\
\Theta &=& \frac{L\theta'_0}{w} \nonumber \\
{\cal L} &=& \frac{L}{\sqrt{1+\Theta^2/2+\Gamma^2/2}}.
\end{eqnarray}
These parameters appear naturally in the below expressions.

To proceed with the analysis, let us assume that the pump field is
a Gaussian beam focused on the crystal so that the Rayleigh range
is much longer than the crystal length:
$\tilde{E}_{\mbox{\scriptsize
P}}(k_x,k_y,\omega)=A_{\mbox{\scriptsize P}}(\omega)
w_{\mbox{\scriptsize P}}\exp\left[-\frac{w_{\mbox{\tiny
P}}^2}{4}\left(k_x^2+k_y^2\right) \right]$, where the pump beam
spectrum is $A_{\mbox{\scriptsize P}} (\omega) = \exp\left(
-\frac{(\omega-2\omega_0)^2} {2\sigma_{\mbox{\tiny P}}^2} \right)$
and the fibers are oriented at the angles for which the degenerate
photons are collected the most efficiently:
$\theta_{\mbox{\scriptsize s}}=\theta_0$ and
$\theta_{\mbox{\scriptsize i}}=-\theta_0$.

The explicit form of the probability amplitude modulus
$|\psi(\omega_{\mbox{\scriptsize s}},\omega_{\mbox{\scriptsize
i}})|$ reads:

\begin{widetext}

\begin{eqnarray}
\label{eq-PsiFrequencies2} |\psi(\omega_{\mbox{\scriptsize
s}},\omega_{\mbox{\scriptsize i}})| &=& \frac{8\pi
w_{\mbox{\scriptsize P}}{\cal L}} {w^2+2w_{\mbox{\scriptsize
P}}^2} \exp\left( -\frac{\Gamma^2}{2+\Gamma^2}
-\frac{1}{2}\boldsymbol{\omega}^T\boldsymbol{\Omega}\,\boldsymbol{\omega}\right)
\times
\nonumber \\
\nonumber \\
& & \exp\left[ -\frac{2(2+\Gamma^2)/w^2}{2+\Theta^2+\Gamma^2}
\left(h-\frac{L\theta'_0}{2}\frac{1+\Gamma^2}{1+\Gamma^2/2}
\right)^2\right],\nonumber \\
\end{eqnarray}
where $\boldsymbol{\omega}=(\omega_{\mbox{\scriptsize
s}}-\omega_0,\omega_{\mbox{\scriptsize i}}-\omega_0)$ and
particular components of the appearing spectrum matrix

\begin{equation}
\boldsymbol{\Omega}= \left(\begin{array}{cc}
\Omega_{\mbox{\scriptsize ss}} & \Omega_{\mbox{\scriptsize si}} \\
\Omega_{\mbox{\scriptsize is}} & \Omega_{\mbox{\scriptsize ii}} \\
\end{array}
\right)
\end{equation}
read:

\begin{eqnarray}
\Omega_{\mbox{\scriptsize ss}} &=& \frac{{\cal L}^2}{8}
\left(\Delta\beta_{z,-}+\frac{\theta_0 \theta'_0}{c}
+\frac{\theta_0
\theta'_0}{\sqrt{2}c}\frac{w\Gamma/\Theta}{\sqrt{w^2+2w_{\mbox{\scriptsize
P}}^2}}\right)^2 + \frac{w^2 w_{\mbox{\scriptsize P}}^2
\theta_0^2}{2c^2(w^2+2w_{\mbox{\scriptsize P}}^2)} +
\frac{1}{\sigma_{\mbox{\scriptsize P}}^2} +
\frac{1}{\sigma_{\mbox{\scriptsize F}}^2}
\nonumber \\
\nonumber \\
\Omega_{\mbox{\scriptsize ii}} &=& \frac{{\cal L}^2}{8}
\left(\Delta\beta_{z,-}+\frac{\theta_0 \theta'_0}{c}
-\frac{\theta_0
\theta'_0}{\sqrt{2}c}\frac{w\Gamma/\Theta}{\sqrt{w^2+2w_{\mbox{\scriptsize
P}}^2}}\right)^2 + \frac{w^2 w_{\mbox{\scriptsize P}}^2
\theta_0^2}{2c^2(w^2+2w_{\mbox{\scriptsize P}}^2)} +
\frac{1}{\sigma_{\mbox{\scriptsize P}}^2} +
\frac{1}{\sigma_{\mbox{\scriptsize F}}^2}
\nonumber \\
\nonumber \\
\Omega_{\mbox{\scriptsize si}} &=& \Omega_{\mbox{\scriptsize is}}
= \frac{{\cal L}^2}{8}
\left(\left(\Delta\beta_{z,-}+\frac{\theta_0
\theta'_0}{c}\right)^2-\left( \frac{\theta_0
\theta'_0}{\sqrt{2}c}\frac{w\Gamma/\Theta}{\sqrt{w^2+2w_{\mbox{\scriptsize
P}}^2}}\right)^2\right) - \frac{w^2 w_{\mbox{\scriptsize P}}^2
\theta_0^2}{2c^2(w^2+2w_{\mbox{\scriptsize P}}^2)} +
\frac{1}{\sigma_{\mbox{\scriptsize P}}^2}.\nonumber \\
\end{eqnarray}
\end{widetext}
Let us notice, that the diagonal elements are not exactly the
same. The symmetry of the coincidence spectrum is broken by the
non-vertical orientation of the crystal's optical axis. However,
for small angles $\theta_0$ the spectrum matrix
$\boldsymbol{\Omega}$ simplifies and the asymmetry of the spectrum
disappears:

\begin{eqnarray}
\label{eq-SpectrumApprox} \Omega_{\mbox{\scriptsize ss}} &\approx&
\Omega_{\mbox{\scriptsize ii}} \approx \frac{{\cal L}^2
\Delta\beta_{z,-}^2}{8} + \frac{w^2 w_{\mbox{\scriptsize P}}^2
\theta_0^2}{2c^2(w^2+2w_{\mbox{\scriptsize P}}^2)}+
\frac{1}{\sigma_{\mbox{\scriptsize P}}^2} +
\frac{1}{\sigma_{\mbox{\scriptsize F}}^2}
\nonumber \\
\nonumber \\
\Omega_{\mbox{\scriptsize si}} &=& \Omega_{\mbox{\scriptsize is}}
\approx \frac{{\cal L}^2 \Delta\beta_{z,-}^2}{8} - \frac{w^2
w_{\mbox{\scriptsize P}}^2
\theta_0^2}{2c^2(w^2+2w_{\mbox{\scriptsize P}}^2)} +
\frac{1}{\sigma_{\mbox{\scriptsize P}}^2}.
\end{eqnarray}
Non-separability of the function $\psi(\omega_{\mbox{\scriptsize
s}},\omega_{\mbox{\scriptsize i}})$ is a signature of frequency
entanglement, a feature that is undesirable if we are interested
in polarization entanglement. To minimize this entanglement one
has to adjust the parameters so that the off-diagonal elements of
the spectrum matrix $\boldsymbol{\Omega}$ become relatively small,
i.e. when $\frac{\Omega_{\mbox{\scriptsize
si}}}{\Omega_{\mbox{\scriptsize ss}}}$ reaches its minimum.
Inspection of expressions (\ref{eq-SpectrumApprox}) shows that for
small angles $\theta_0$ and small mode waists $w$ and
$w_{\mbox{\scriptsize P}}$ the separability can be accomplished
only with the use of narrow-band interference filters. However,
for large mode waists one can separate the spectrum by fulfilling
the following condition:

\begin{equation}
\frac{w^2 w_{\mbox{\scriptsize P}}^2
\theta_0^2}{2c^2(w^2+2w_{\mbox{\scriptsize P}}^2)} -\frac{{\cal
L}^2 \Delta\beta_{z,-}^2}{8} = \frac{1}{\sigma_{\mbox{\scriptsize
P}}^2}.
\end{equation}
Notice that this condition does not depend on the interference
filters' bandwidth $\sigma_{\mbox{\scriptsize F}}$.

\begin{figure}
\begin{center}
\epsfig{file=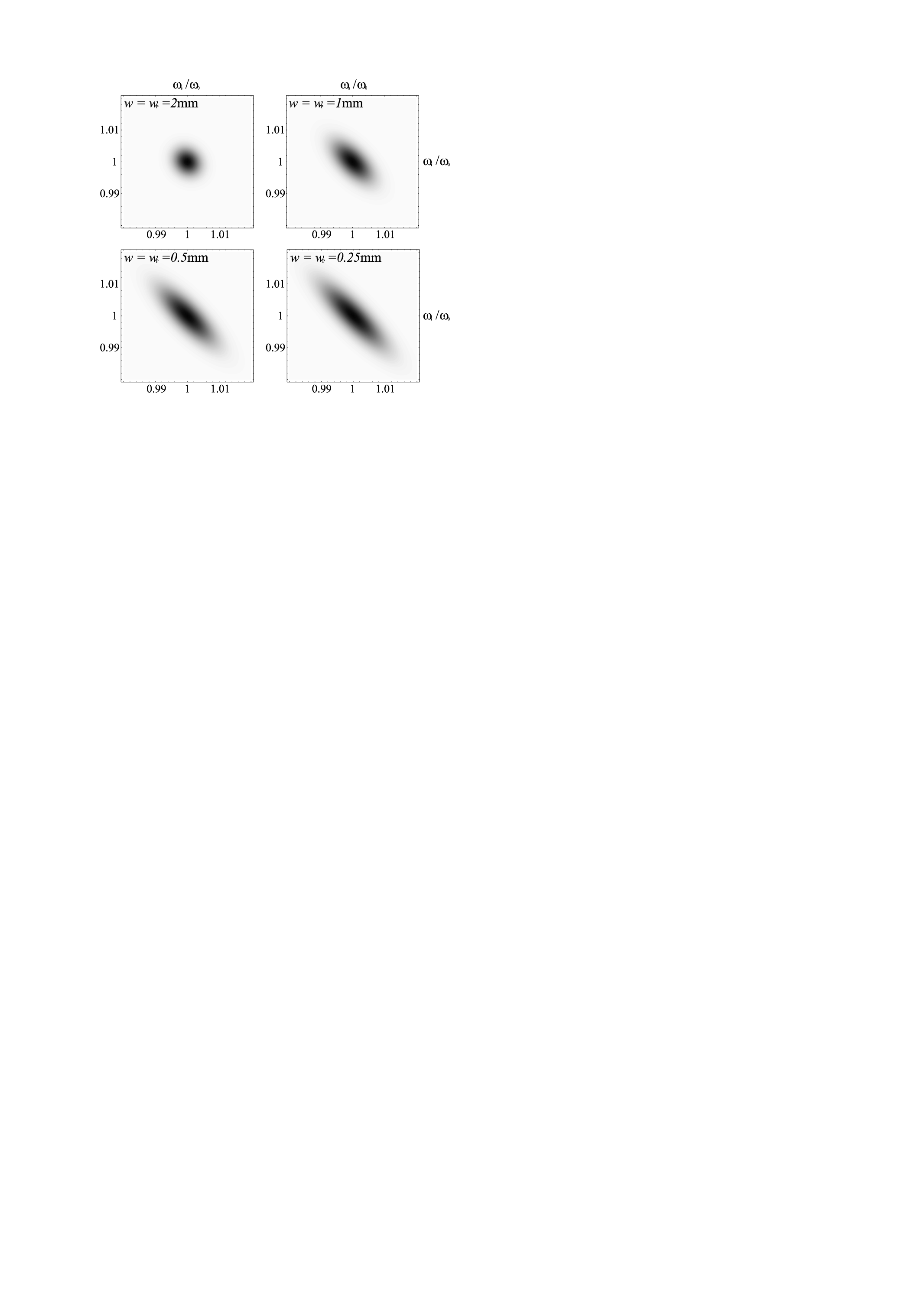}
\end{center}
\caption{\label{fig-widma} Coincidence spectrum
$|\psi(\omega_{\mbox{\scriptsize s}},\omega_{\mbox{\scriptsize
i}})|^2$ for several mode waist diameters. For strongly focused
beams, the spectrum becomes inseparable, and the separability can
be achieved only for unfocused beams.}
\end{figure}

For concreteness, we have analyzed the coincidence spectrum
(\ref{eq-PsiFrequencies2}) on the example of a BBO crystal cut for
the degenerate frequency $\omega_0 = 780\mbox{nm}$ and the
emission angle $\theta_0=1.4^\circ$ (using the Sellmeier formulas
for the indices of refraction \cite{ref-ZielonaKsiazka} one can
calculate that in this case $\Delta\beta_{-,z}=2.06\cdot 10^6
\frac{\mbox{s}}{\mbox{m}}$), pump beam spectrum width $5\mbox{nm}$
to and the interference filter's width $17\mbox{nm}$. The results
are shown on Figure \ref{fig-widma} for several mode waist
diameters. We see that when the beams are strongly focused, the
spectrum $|\psi(\omega_{\mbox{\scriptsize
s}},\omega_{\mbox{\scriptsize i}})|^2$ becomes inseparable, and
the separability can be achieved only for unfocused beams. In the
considered case the optimum waists are approximately
$w=w_{\mbox{\scriptsize P}}\approx 2.5\mbox{mm}$.

Expression (\ref{eq-PsiFrequencies2}) can be used to maximize the
coincidence probability (\ref{eq-PsiFrequencies}) over all
possible choices of the parameter $h$. The optimal shift $h$ of
the coupled beam in respect to the pumping beam at $z=z_2$ is:

\begin{equation}
h=\frac{L\theta'_0}{2}\frac{1+\Gamma^2}{1+\Gamma^2/2}.
\end{equation}
One can notice that in the absence of the walk-off effect
affecting the propagation of the pumping beam (for $\Gamma=0$),
the optimal choice of $h$ involves crossing of the pumping mode
and two fiber coupled modes exactly in the middle of the nonlinear
medium. The modification comes from the fact that the pumping beam
is transversely shifted due to the walk-off effect during the
propagation in the crystal. The effects of the walk-off become
more significant when all the modes are strongly focused on the
crystal, because only the relative shift of the beam in comparison
to its waist and the waist of the coupled modes is relevant.

To obtain the total probability $p$ of coupling any photon pair
into the two fibers we need to integrate the probability
$|\psi(\omega_{\mbox{\scriptsize s}},\omega_{\mbox{\scriptsize
i}})|^2$ over all possible frequencies. Notice that this time we
integrate the probabilities, not their amplitudes as the photons
of different frequencies are perfectly distinguishable:

\begin{equation}
p=\int\mbox{d}\omega_{\mbox{\scriptsize s}}
\mbox{d}\omega_{\mbox{\scriptsize i}}
|\psi(\omega_{\mbox{\scriptsize s}},\omega_{\mbox{\scriptsize
i}})|^2.
\end{equation}
The above integration can be also done analytically if we extend
the lower limits to $-\infty$. In this case, for the optimum
choice of $h$ we obtain a relatively simple expression for the
probability $p$:

\begin{equation}
\label{eq-Probability} p = \frac{64\pi^3 w_{\mbox{\scriptsize
P}}^2{\cal L}^2}{(w^2+2w_{\mbox{\scriptsize
P}}^2)^2\sqrt{\det\boldsymbol{\Omega}}}\exp\left(
-\frac{2\Gamma^2}{2+\Gamma^2} \right).
\end{equation}
Before we continue with a detailed analysis of the above result,
let us stress once more, that it is valid only when the Rayleigh
range of all the considered beams is much longer than the
thickness of the crystal, i.e. for $\frac{1}{2}k_0w^2\gg L$, where
$k_0$ is a characteristic wave-vector. This condition guarantees
that the diffraction of the pump beam and generated photon pairs
inside the crystal can be neglected.

Several conclusions about the efficiency of the coupling process
can be drawn from the expression (\ref{eq-Probability}). First and
the most obvious is that the overall probability $p$ does not
depend on $\Delta\beta_{+,z}$. Coupling efficiency is also a
decreasing function of the angle $\theta_0$ (and proportional
$\theta'_0$), as it appears only in the denominator of the
expression (\ref{eq-Probability}). This can be easily understood,
since increasing $\theta_0$ (depending on the geometry of the
crystal and its indices of refraction) decreases the spatial
volume of an overlap between the pumping mode and coupled modes.
Another clear conclusion is that the walk-off effect represented
by $\Gamma$ and responsible for the transverse shift of the pump
beam during the propagation through the crystal, also diminishes
the coupling probability. The next relevant parameter, inverse
group velocity difference of the extraordinary and ordinary beams
propagating towards $z$: $\Delta\beta_{-,z}$ suppresses the
down-conversion process. This effect is also straightforward, as
$\Delta\beta_{-,z}$ is inversely proportional to the
down-conversion spectrum width, so the larger $\Delta\beta_{-,z}$
is, the fewer frequencies are efficiently down-converted.

\begin{figure}
\begin{center}
\epsfig{file=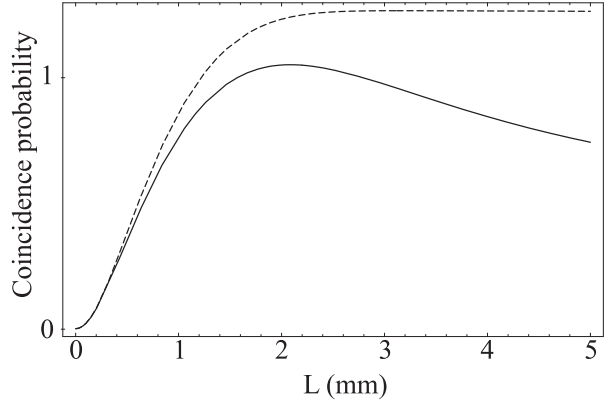}
\end{center}
\caption{\label{fig-PeOdEl} Relative coincidence probability as a
function of the crystal length $L$ for fixed
$w=w_{\mbox{\scriptsize P}}=100\mu\mbox{m}$. Solid line -
analytical approximation, dashed line - numerical result. The
Rayleigh range for the pumping beam is equal to $13\mbox{cm}$.}
\end{figure}

The next issue of practical interest is how the probability $p$
depends on the parameters adjustable for a given nonlinear
material, namely the crystal length $L$ and the mode waists $w$
and $w_{\mbox{\scriptsize P}}$. Again, we have analyzed this issue
on the same example of the BBO crystal. For such setting we have
calculated the pair coupling probability $p$ given by
Eq.~(\ref{eq-Probability}) and compared it with the result of a
numerical calculation. In the figure \ref{fig-PeOdEl} we show the
probability $p$ as a function of the crystal length $L$ for fixed
$w=w_{\mbox{\scriptsize P}}=100\mu\mbox{m}$ (solid line) and an
analogous numerical result (dashed line). The only approximation
that the numerical calculus involves is the expansion of the
$\delta_\pm$ functions to the second order of the Taylor series.
In the numerical case we keep the $\mbox{sinc}\,x$ function
unchanged, as well as we do not assume that the Rayleigh range of
the considered modes exceeds the crystal length.

It is clear from the Figure \ref{fig-PeOdEl}, that the analytic
approximation breaks down when we consider thick crystals. This is
when the argument of the $\mbox{sinc}\,x$ function becomes large,
the wave tails of the $\mbox{sinc}\,x$ cease to be insignificant -
and the Gaussian approximation is not legitimate anymore. However
for thin crystals all the analytic approximations work well. We
will show later that this region is the most interesting when one
intends to maximize the coupling efficiency.

\begin{figure}
\begin{center}
\epsfig{file=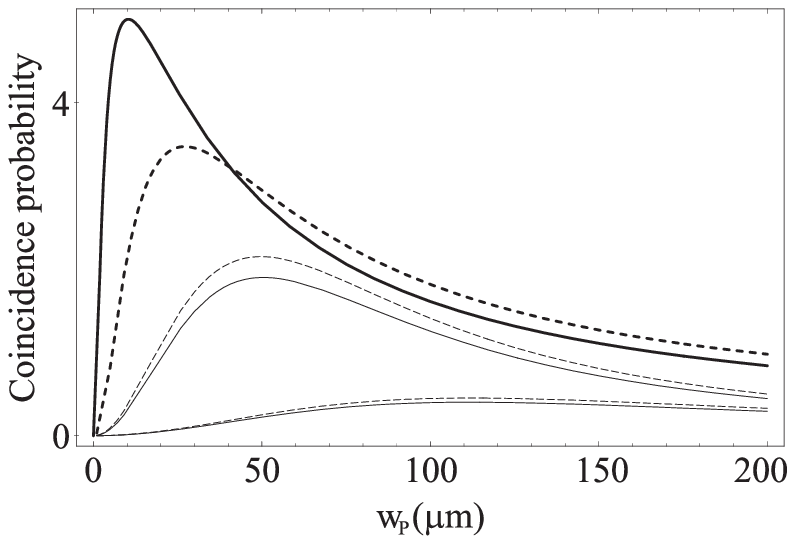}
\end{center}
\caption{\label{fig-PeOdWuPe} Relative coincidence probability as
a function of the pump beam waist $w_{\mbox{\scriptsize P}}$:
solid line - analytical approximation, dashed line - numerical
result for several cases. The bottom curves- for $L=1\mbox{mm}$
and $w=50\mu\mbox{m}$, the middle curves - for $L=1\mbox{mm}$ and
$w=150\mu\mbox{m}$ and the top, bold curves for $L$ and $w$ chosen
to maximize the result for each $w_{\mbox{\scriptsize P}}$.}
\end{figure}

In the Figure \ref{fig-PeOdWuPe} we have shown the coincidence
probability as a function of the pump beam waist
$w_{\mbox{\scriptsize P}}$. Solid lines again represent the
results of the analytical approximation (\ref{eq-Probability}),
dashed lines - the analogous numerical results. The plots
demonstrate several cases: the bottom curves - for $L=1\mbox{mm}$
and $w=50\mu\mbox{m}$, the middle curves - for $L=1\mbox{mm}$ and
$w=150\mu\mbox{m}$ and the top, bold curves show the probability
$p$ as a function of $w_{\mbox{\scriptsize P}}$ with optimized
values of $L$ and $w$.

The figures reveal a good agreement between the analytical theory
and the numerical calculations. The discrepancies are more
relevant only in the region of thick crystals or a very intense
focusing of the beams.

To find the optimum choice of $L$ and $w$ we have numerically
maximized the probability $p$ given by the expression
(\ref{eq-Probability}) in the region depicted in Figure
\ref{fig-PeOdWuPe} and performed a linear fit to the obtained
dependencies. As a result we have obtained:

\begin{eqnarray}
\label{eq-optimum}
w_{\mbox{\scriptsize opt}} &\approx& 3.32 \mu\mbox{m} + 0.308 w_{\mbox{\scriptsize P}} \nonumber \\
L_{\mbox{\scriptsize opt}} &\approx& 167 \mu\mbox{m} + 14.8 w_{\mbox{\scriptsize P}}. \nonumber \\
\end{eqnarray}
The bolder top curves on the Figure \ref{fig-PeOdWuPe} (analytical
and numerical) involve an application of the above dependencies.

From the Figure \ref{fig-PeOdWuPe}, one can notice that the
coupling process is the most effective for a very small pump beam
mode waist. According to the numerical result depicted in this
figure we find that the optimum waist $w_{\mbox{\scriptsize P,
opt}} \approx 27\mu\mbox{m}$ which, according to the expressions
(\ref{eq-optimum}), corresponds to $w_{\mbox{\scriptsize opt}} =
12\mu\mbox{m}$ and $L_{\mbox{\scriptsize opt}} = 0.57\mbox{mm}$.
Rayleigh range for the optimally focused pumping beam equals
$98\mbox{mm}$. Therefore the condition of the large Rayleigh range
in comparison with the crystal length is well preserved.

\section{Double crystal\label{sec-Double}}
So far we have considered the case of a single nonlinear crystal.
However, to generate polarization entanglement with the use of
type-I process in the scheme proposed by Kwiat {\em et al.}
\cite{ref-Kwiat1999} one needs two crystals rotated in respect to
each other by the angle of $90^\circ$ around the $z$ axis. In this
scheme a pump beam is linearly polarized so that the chance for a
down-conversion in the first and in the second crystal are exactly
the same.

If the first crystal is oriented as discussed in the previous
sections (the optical axis of the crystal lying in a plane
oriented at the angle $45^\circ$ in respect to the horizontal
plane), then we have two options concerning the orientation of the
second crystal. It may be either horizontally or vertically
flipped in respect to the geometry of the first crystal. Let us
notice that in the latter case, when the crystals are symmetric in
respect to the horizontal plane, the amplitude
$\psi(\omega_{\mbox{\scriptsize s}}, \omega_{\mbox{\scriptsize
i}})$ given by equation (\ref{eq-PsiFrequencies}) is exactly the
same for both crystals. Therefore the distinguishability of the
photon pairs from the both crystals may only be a consequence of
either spatial separation of the emission cones or propagation
effects: the photons from the first crystal propagate through the
second one as the extraordinary rays and undergo a transversal
shift due to the walk-off effect. Secondly, the pump beam
polarization component down-converting in the second crystal and
propagating through the first one is delayed in respect to the
other polarization component. Consequently, photons generated in
both crystals arrive to the fibers at different times. However,
these two dominant propagation effects are related to linear phase
shifts easily compensable with additional linear optics. The pump
beam delay may be compensated by a birefringent material inserted
in front of the double crystal, and the walk-off effect can be
compensated by a pair of inverted down-conversion crystals placed
in front of the fibers.

At this point our motivation concerning the diagonal orientation
of the crystals becomes clear. In the proposed configuration only
the spatial separation of the photon emission cones can give rise
to the photon distinguishability. However for thin crystals
separated by a small distance, one can attain the perfect
indistinguishability of photon pairs emerging from the both
crystals and consequently maximize the polarization entanglement.

\section{Conclusions\label{sec-Conclusions}}
We have developed the approximate analytic theory of a type-I
down-conversion to analyze the properties of the fiber coupled
photon pairs. We have shown that in the regime of the weak
focusing one can separate the coincidence spectrum, while for
strong focusing one can optimize the coincidence efficiency. We
have analytically derived the coincidence spectrum separability
condition and the pair fiber-coupling probability and compared the
result with the numerical calculation obtaining a good agreement.

\section*{Acknowledgments}
I would like to thank K. Banaszek for inspiration and help with
this work and C. Radzewicz, W. Wasilewski and K. W\'{o}dkiewicz
for many valuable discussions and comments on the manuscript. This
work has been supported by the Polish Committee for Scientific
Research (KBN Grant no. PBZ/KBN/043/P03/2001) and is a part of the
general program on quantum engineering of the FAMO National
Laboratory in Toru\'{n}, Poland. I also thank The Foundation for
Polish Science for the support with the Annual Stipend for Young
Scientists.

\appendix
\section*{Appendix}
Let us introduce a following coordinate system: $Z$ axis
perpendicular to the crystal and transverse coordinates' axes $X$
and $Y$, so that the optical axis lies in the plane $XZ$, at the
angle $\alpha$ respect to the $Z$ axis. The dispersion relations
for the ordinary and extraordinary beams have the form,
respectively \cite{ref-Yariv}:

\begin{widetext}
\begin{eqnarray}
k^{\mbox{\scriptsize o}}_Z(k_X,k_Y,\omega) &=&
\sqrt{\frac{\omega^2 n^{\mbox{\scriptsize
o}}(\omega)^2}{c^2}-k_X^2-k_Y^2}
\nonumber \\ \nonumber \\
k^{\mbox{\scriptsize e}}_Z(k_X,k_Y,\omega) &=& \frac{k_X
\sin\alpha\cos\alpha\left(1-\frac{n^{\mbox{\tiny e}}(\omega)^2}
{n^{\mbox{\tiny o}}(\omega)^2}\right)}
{\sin^2\alpha+\frac{n^{\mbox{\tiny e}}(\omega)^2} {n^{\mbox{\tiny
o}}(\omega)^2}\cos^2\alpha}+
\nonumber \\ \nonumber \\
& &\frac{\sqrt{\left( \frac{\omega^2 n^{\mbox{\scriptsize
e}}(\omega)^2}{c^2}-k_Y^2 \right)
\left(\sin^2\alpha+\frac{n^{\mbox{\tiny e}}(\omega)^2}
{n^{\mbox{\tiny
o}}(\omega)^2}\cos^2\alpha\right)-k_X^2\frac{n^{\mbox{\tiny
e}}(\omega)^2} {n^{\mbox{\tiny o}}(\omega)^2}}}
{\sin^2\alpha+\frac{n^{\mbox{\tiny e}}(\omega)^2} {n^{\mbox{\tiny
o}}(\omega)^2}\cos^2\alpha},
\end{eqnarray}
\end{widetext}
where $n^{\mbox{\scriptsize o}}(\omega)$ and $n^{\mbox{\scriptsize
e}}(\omega)$ are the ordinary and extraordinary indices of
refraction, respectively, given by the Sellmeier formulas
\cite{ref-ZielonaKsiazka}.

From the above relations one can calculate the quantities
appearing in Equation (\ref{eq-mismatchderivatives}): walk-off
angle $\gamma$, group velocity combinations $\Delta\beta_{\pm,z}$
and angle of propagation of the degenerate photons inside the
crystal $\theta'_0$ defined as:

\begin{eqnarray}
\gamma &=& \frac{\partial k^{\mbox{\scriptsize
e}}_Z(0,0,2\omega_0)}{\partial
k_X} \nonumber \\ \nonumber \\
\Delta\beta_{\pm,z} &=& \frac{\partial k^{\mbox{\scriptsize e}}_Z
(0,0,2\omega_0)} {\partial \omega} \pm \frac{\partial
k^{\mbox{\scriptsize o}}_Z(0,0,\omega_0)} {\partial \omega} \nonumber \\ \nonumber \\
\theta'_0 &=& \frac{\theta_0}{n^{\mbox{\tiny o}}(\omega_0)}.
\end{eqnarray}

\newpage

~

\newpage

~

\newpage

~

\newpage

\begin{thebibliography}{99}

\bibitem{ref-Kwiat1995}
P. G. Kwiat, K. Mattle, H. Weinfurter, A. Zeilinger, A. V.
Sergienko, and Y. Shih Phys. Rev. Lett. {\bf 75}, 4337 (1995).

\bibitem{ref-Monken1998}
C. H. Monken, P. H. Souto Ribeiro, and S. Pádua Phys. Rev. A {\bf
57}, R2267 (1998).

\bibitem{ref-Kurtsiefer2001}
C. Kurtsiefer, M. Oberparleiter and H. Weinfurter, Phys. Rev. A
{\bf 64}, 023802 (2001).

\bibitem{ref-Bovino2003}
F. A. Bovino, P. Varisco, A. M. Colla, G. Castagnoli, G. Di
Giuseppe and A. V. Sergienko, quant-ph/0303126 (2003).

\bibitem{ref-Castalletto2003}
S. Castelletto, I. P. Degiovanni, M. Ware and A. Migdall,
quant-ph/0311099 (2003).

\bibitem{ref-Banaszek2004}
K. Banaszek, A. Dragan, W. Wasilewski and C. Radzewicz, Phys. Rev.
Lett. (in press).

\bibitem{ref-URen2003}
A. B. U'Ren, K. Banaszek and I. Walmsley, Quant. Inf. Comp. {\bf
3}, 480 (2003).

\bibitem{ref-Keller1997}
T. E. Keller and M. H. Rubin Phys. Rev. A {\bf 56}, 1534 (1997).

\bibitem{ref-Rubin1997}
M. H. Rubin Phys. Rev. A {\bf 54}, 5349 (1996).

\bibitem{ref-Yariv}
A. Yariv, {\em Quantum Electronics}, 3rd edition, Wiley, New York
(1988).

\bibitem{ref-Klyshko1988}
D. N. Klyshko, {\em Photons and Nonlinear Optics}, Gordon and
Breach Science Publish., New York (1988).

\bibitem{ref-ZielonaKsiazka}
D. N. Nikogosyan, {\em Poperties of optical and laser-related
materials}, Wiley, New York (1997).

\bibitem{ref-Kwiat1999}
P. G. Kwiat, E. Waks, A. G. White, I. Appelbaum and P. H.
Eberhard, Phys. Rev. A {\bf 60}, R773 (1999).





\end{thebibliography}
\end{document}